\documentclass[prd,aps, tightenlines, preprintnumbers, showpacs, nofootinbib,superscriptaddress,notitlepage]{revtex4-1}

\usepackage{pgfplots}
\usepackage{filecontents}
\usepackage{standalone}

\usepackage{graphicx}
\usepackage{amsmath,amsthm,amssymb}
\usepackage{color}

\newcommand{\non}{\nonumber\\}

\usepackage[normalem]{ulem}

\begin{document}
\title{{Nuclear modification of forward $D$ production in pPb collisions at the LHC} 
}

\author{Hirotsugu Fujii}
\affiliation{Institute of Physics, University of Tokyo, Tokyo 153-8902, Japan}

\author{Kazuhiro Watanabe}
\email{Contact for numerical data : \href{watanabe@jlab.org}{watanabe@jlab.org}}
\affiliation{Physics Department, Old Dominion University, Norfolk, VA 23529, USA}
\affiliation{Theory Center, Thomas Jefferson National Accelerator Facility, Newport News, VA 23606, USA}

\preprint{JLAB-THY-17-2506}
\date{\today}

\begin{abstract}
  We study nuclear modification factors for single $D$ meson and semileptonic decay lepton $l$ ($=e,\mu$) production in minimum bias proton-nucleus (p$A$) collisions at the LHC in the color-glass-condensate (CGC) framework at leading order in strong coupling. In our numerical computations, transverse momentum ($k_\perp$) dependent multi-point Wilson line correlators are employed for describing target nucleus for p$A$ and proton for pp. The projectile proton is treated with unintegrated gluon distribution function, which is also $k_\perp$-dependent.  The rapidity evolutions of these functions in the small Bjorken $x$ region are taken into account by solving running coupling Balitsky-Kovchegov (BK) equation at leading logarithmic accuracy. For simplicity, we employ Kartvelishvili's type fragmentation function and a simple model for lepton energy distribution from seileptonic decay, respectively, to compute differential cross sections for $D$ and $l$ production. The gluon saturation scale inside the heavy nucleus is enhanced and dependent on $x$, which we take into account by replacing the initial saturation scale in the BK equation with a larger value for the heavy nucleus. We show that the saturation effect leads to perceptible nuclear suppression of $D$ production at forward rapidity. Our numerical results predict similar nuclear suppressions in p$A$ collisions for forward $l$ production at lower transverse momentum $p_\perp<2\;{\rm GeV}$. Numerical tables on the nuclear modifications of $D$ and $l$ are listed in this note.
\end{abstract}

\maketitle


\section{Background}{\label{section:background}}

Exploring gluon saturation phenomenon inside the hadron has been an active research subject in nuclear physics. Coherence of gluons inside the hadron, which is responsible for the gluon saturation, stems from a dynamical competition between gluon bremsstrahlung and recombination at small values of Bjorken $x$~\cite{Gribov:1984tu,Mueller:1985wy}. Importantly, the saturation dynamics becomes more noticeable when the heavier nucleus is taken as a scattering target instead of the proton, because the saturation scale $Q_s$ is enhanced by the coherence along the target thickness which is in proportion to $A^{1/3}$ with $A$ atomic mass number. In order to describe this small-$x$ part of the hadron and/or nuclear wavefunction incorporating the gluon saturation dynamics, the color-glass-condensate (CGC) framework has been elaborated over the last two decades~\cite{Gelis:2010nm,Kovchegov:2012mbw}. 

At present, the CERN Large Hadron Collider (LHC) is one of the most powerful facilities and it provides us with the most energetic p$A$ collisions as a good testing ground for the gluon saturation phenomenon. It is customarily assumed that secondary scatterings among the produced particles are less important in p$A$ collisions, in contrast to heavy nucleus-nucleus ($AA$) collisions~\footnote{Recently, anisotropic momentum distributions of light hadrons are observed in p$A$ and even in pp collisions~\cite{Abelev:2012ola,Abelev:2014mda,Aad:2012gla,Aad:2013fja,Aad:2014lta,CMS:2012qk,Chatrchyan:2013nka,Adare:2013piz,Adare:2014keg,Khachatryan:2016txc}, whose origin is now under discussion in the field.}. Therefore p$A$ collision experiment is extremely beneficial to pin down the onset of the gluon saturation inside nucleus.

Observables for probing deep saturation regime in the wavefunction of heavy nuclei in high energy p$A$ collisions include production of quarkonia/open heavy-flavor mesons at forward rapidity. These are gripping probes because the saturation scale inside the heavy nucleus becomes larger than the heavy quark mass scale. Heavy-flavor production in p$A$ collisions have been also studied for quantifying initial-state nuclear effects, the so-called cold nuclear matter (CNM) effects~\cite{Andronic:2015wma}. Precise calibration of the observables taking account of the CNM effects is very essential, because in high energy $AA$ collisions heavy-flavor production have been considered as invaluable external probes to examine properties of hot QCD matter.

Over the last few years, we have performed numerical calculations of quarkonium production~\cite{Fujii:2013gxa}, open heavy-flavor meson production~\cite{Fujii:2013yja}, and its decay lepton production~\cite{Fujii:2015lld} in p$A$ collisions in the CGC framework. In the meantime, data on quarkonium production in p$A$/d$A$ collisions have been accumulated and reported by RHIC and LHC experiments. Our prediction in the CGC framework showed a discrepancy with the data, and many theoretical attempts have been actively and carefully conducted to understand the LHC data more quantitatively~\cite{Watanabe:2016ert}. On the other hand, unlike quarkonium production which involves bound-state formation dynamics at late stage, production of open heavy-flavor mesons may be treated with the fragmentation functions. Therefore, the initial-state gluon saturation effect can be investigated with less ambiguity. Although data analyses about open heavy-flavor production at the LHC, especially data at forward rapidity, have been ongoing yet, ALICE collaboration reported their result on semileptonic decay $\mu$ production at forward rapidity in pPb collisions~\cite{Acharya:2017hdv}. For $D$ meson production, preliminary but intriguing data by LHCb collaboration are available on-line~\cite{LHCb:2016huj}.

In Refs.~\cite{Fujii:2013yja} and \cite{Fujii:2015lld}, we pointed out that heavy-flavor meson $D$ and even its decay lepton $l=e,\mu$ produced in p$A$ collisions at the LHC in the forward rapidity region carry important information on the gluons lying down in the saturation regime inside heavy nucleus. This implies that we may glimpse the saturation effect by data comparison with theory calculations. Therefore, now, it is very meaningful and important to prepare for comparisons between up-to-date numerical results in the CGC framework and the data currently available as well as upcoming, about heavy-flavor production in p$A$ collisions at the LHC. This note is aimed at providing such comparisons.

This short note is organized as follows: In section~\ref{section:Framework}, the CGC framework for heavy quark production in p$A$ collisions and the description of the fragmentation part for producing $D$ and $l$ are outlined. The differential cross section for single heavy-flavor meson $D$ and lepton $l$ production in the LO CGC framework can be written schematically as 
\begin{align}
d\sigma_{D}&=D_{c}^{D}\otimes \varphi_{\rm p}\otimes \phi_{A}\otimes \Xi_{c},\\
d\sigma_{l}&=\mathcal{F}_{D}^{l}\otimes D_{c}^{D}\otimes \varphi_{\rm p}\otimes \phi_{A}\otimes \Xi_{c}
,
\end{align}
where $\Xi_{c}$ represents short distance hard scattering part shown in Eq.~(\ref{eq:xsection-kt-factorization-LN}), $\varphi_{\rm p}$ is the unintegrated gluon distribution function of the projectile proton in Eq.~(\ref{eq:phip}), and $\phi_{A}$ is the multi-point Wilson line correlator in the target given in Eq.~(\ref{eq:phiA}). $D_{c}^{D}$ and $\mathcal{F}_{D}^{l}$ are the fragmentation function (\ref{eq:fragmentation}) and the lepton decay function (\ref{eq:semileptonic}), respectively. $\otimes$ represents the relevant convolution integrals including constant factors. In our calculations, we take into account the rapidity or energy evolution effects which are embodied through the rapidity dependence of $\varphi_{\rm p}$ and $\phi_{A}$ using the running coupling Balitsky-Kovchegov equation (\ref{eq:bk}) only. Then, the numerical results are presented in section~\ref{section:numerical-results}.

\section{Framework}{\label{section:Framework}}

\subsection{Heavy quark production}

We briefly recapitulate the CGC formula for heavy quark production in p$A$ collisions,
${\rm p}+A\rightarrow Q(q)+\bar{Q}(p)+X$,
where $q$ and $p$ are four momenta of a produced quark ($Q$) and antiquark ($\bar{Q}$), respectively.
In this note, we use the CGC formula which is formally valid only
at leading-order (LO) in strong coupling constant, $\alpha_s$, but
fully takes account of the saturation effects.
We treat only the gluon fusion processes in the heavy $Q\bar Q$ pair production
because the gluon density inside the hadron/nucleus becomes much larger
than the quark density at small $x$ probed in high-energy scatterings.
The gluon distribution inside the target hadron is described with the transverse momentum ($k_\perp$) dependent functions, in contrast to the case of collinear factorization framework where $k_\perp$-dependence is suppressed. The differential cross section for producing a $Q\bar Q$ pair in p$A$ collisions can be written as~\cite{Blaizot:2004wu,Blaizot:2004wv,Fujii:2006ab} (see also \cite{Tuchin:2004rb,Kovchegov:2006qn})
\begin{align}
\frac{d \sigma_{q \bar{q}}}{d^2p_{\perp} d^2q_{\perp} dy_p dy_{q}}
=
\frac{\alpha_s^2}{64\pi^6 C_F}
\int\frac{d^2k_{2\perp}d^2k_\perp}{(2\pi)^4}
\frac{\Xi({k}_{1\perp}, {k}_{2\perp},{k}_{\perp})}
{k_{1\perp}^2 k_{2\perp}^2}
\; 
\varphi_{{\rm p},x_1}(k_{1\perp})
\; 
\phi_{A,x_2}({k}_{2\perp},{k}_\perp),
\label{eq:xsection-kt-factorization-LN}
\end{align}
where the large-$N_c$ approximation has been assumed. 
$\Xi$ is the partonic hard scattering part and can be decomposed into $\Xi=\Xi^{q\bar{q},q\bar{q}}+\Xi^{q\bar{q},g}+\Xi^{g,g}\;$ with
\begin{align}
&\Xi^{q\bar{q},q\bar{q}}
=\;\frac{32 p^+q^+(m^2+a_\perp^2)(m^2+b_\perp^2)}{[2p^+(m^2+a_\perp^2)+2q^+(m^2+b_\perp^2)]^2},
\label{eq:xiqq}\\
&\Xi^{q\bar{q},g}
=\;\frac{16}{2(m^2+p\cdot q)[2p^+(m^2+a_\perp^2)+2q^+(m^2+b_\perp^2)]}
\Bigg[(m^2+a_\perp\cdot b_\perp)\left\{q^+C\cdot p+p^+C\cdot q-C^+(m^2+p\cdot q)\right\}\non
&+C^+\left\{(m^2+b_\perp\cdot q_\perp)(m^2-a_\perp\cdot p_\perp)-(m^2+a_\perp\cdot q_\perp)(m^2-b_\perp\cdot p_\perp)\right\}\non
&+p^+\left\{a_\perp\cdot C_\perp(m^2+b_\perp\cdot q_\perp)-b_\perp\cdot C_\perp(m^2+a_\perp\cdot q_\perp)\right\}+q^+\left\{a_\perp\cdot C_\perp(m^2-b_\perp\cdot p_\perp)-b_\perp\cdot C_\perp(m^2-a_\perp\cdot p_\perp)\right\}\Bigg],
\label{eq:xiqg}\\
&\Xi^{g,g}
=\;\frac{4\left[2(p\cdot C)(q\cdot C)-(m^2+p\cdot q)C^2\right]}{4(m^2+p\cdot q)^2},
\label{eq:xigg}
\end{align}
where $a_\perp=q_\perp-k_\perp$, $b_\perp=q_\perp-k_\perp-k_{1\perp}$ and 
$C^+=\;p^++q^+-\frac{k_{1\perp}^2}{p^-+q^-}$,~$C^-=\;\frac{k_{2\perp}^2}{p^++q^+}-(p^-+q^-)$, and~$C_\perp=\;k_{2\perp}-k_{1\perp}$. 
$x_{1,2}$ are longitudinal momentum fractions of the gluons from the projectile proton and target nucleus, respectively. From $2\rightarrow2$ kinematics, measurement of the produced $Q\bar Q$ pair probes $x_{1,2} = (m_{q\perp}e^{\pm y_{q}}+m_{p\perp}e^{\pm y_{p}})/\sqrt{s}$ with transverse mass $m_{q\perp}=\sqrt{m^2+q_{\perp}^2}$.
$\varphi_{{\rm p},x_1}(k_{1\perp})$ is the unintegrated gluon distribution function (UGDF) of the projectile proton, and $\phi_{{A},x_2}({k}_{2\perp},{k}_\perp)$ is the multi-point Wilson line correlator for the target nucleus in the large-$N_c$ limit, which are given as follows:
\begin{align}
\varphi_{{\rm p},x}(k_{1\perp}) &= S_{{\rm p}\perp}\, \frac{N_ck_{1\perp}^2}{4\alpha_s} \int\frac{d^2l_\perp}{(2\pi)^2}F_{x}({k}_{1\perp}-{l}_\perp)F_{x}(l_{\perp}),
\label{eq:phip}
\\
\phi_{{A},x}({k}_{2\perp},{k}_\perp)
&=S_{A\perp} \,\frac{N_c k_{2\perp}^2}{4\alpha_s}F_{x}({k}_{2\perp}-{k}_\perp)F_{x}(k_\perp).
\label{eq:phiA}
\end{align}
$S_{{\rm p}\perp}$ ($S_{A\perp}$) is effective transverse area of the proton (nucleus). In these formulas, we have assumed translational invariance in the transverse plane of the projectile and target, to factorize $S_{{\rm p}\perp}$  and $S_{A\perp}$ explicitly. The Fourier transform of the fundamental dipole amplitude is defined as ${F}_{x}(k_{\perp})\equiv\int d^2x_\perp e^{-ik_{\perp}\cdot x_\perp} S_x(x_\perp)=\int d^2x_\perp e^{-ik_{\perp}\cdot x_\perp} \left<{\mathrm Tr}\left[U(x_\perp)U^\dagger(0_\perp)\right]\right>_x/N_c$. $U(x_\perp)$ is the fundamental Wilson line which represents multiple interactions with the gauge fields and a resultant gauge rotation occurring when a high-energy quark traverses through the background gluon fields at a transverse position $x_\perp$. The differential cross section for single $Q$ production is obtained by integrating the pair cross-section (\ref{eq:xsection-kt-factorization-LN}) over the phase space of $\bar Q$:
\begin{align}
\frac{d\sigma_{Q}}{dq_\perp dy_q}=\int d^2p_\perp dy_p \frac{d\sigma_{Q\bar Q}}{d^2q_\perp d^2p_\perp dy_pdy_q}.
\end{align}

\subsection{Fragmentation part}

We adopt a heavy-quark fragmentation function for describing open heavy-flavor meson production. Hereinafter, we restrict our discussion to $D$ meson production. Production of $D$ meson ($D^0$, $D^+$, $D^{\ast+}$, etc.) from charm quark $c$ is described with the heavy-quark fragmentation function $D_c^D(z)$ as 
\begin{align}
\frac{d \sigma_{D}}{d^2p_{D\perp} dy}
=
Br(c\rightarrow D)
\int dz\frac{D_c^D(z)}{z^2}
\frac{d \sigma_{c}}{d^2p_{c\perp} dy}
\, ,
\label{eq:fragmentation}
\end{align}
where the rapidity is set to $y\equiv y_c=y_{D}$ and the momentum fraction $z$ is defined by $p_{D\perp}\equiv zp_{c\perp}$. We use  the fragmentation function of Kartvelishvili's form~\cite{Kartvelishvili:1977pi}: $D_c^D(z)=(\alpha+1)(\alpha+2) z^{\alpha}(1-z)$ with a parameter $\alpha=3.5$ for $D$ meson. We have confirmed that use of different parametrization of $D_c^D(z)$ such as Peterson's type fragmentation function~\cite{Peterson:1982ak} does not make a substantial change of $p_{D\perp}$ distribution of nuclear modification factor which we will discuss later. $Br(c\rightarrow D)$ is the branching ratio for the transition probability from $c$ to $D$, and satisfies $\sum_X Br(c\rightarrow X)=1$ with $X$ being possible heavy-flavor hadrons. Indeed, the branching ratio is not much important as long as we deal with the nuclear modification factor, since it is likely to cancel out in the ratio between the cross sections of pp and p$A$ collisions.

Next, for describing semileptonic decay processes (e.g. $D \to K l \bar \nu$), we introduce the lepton decay function ${\cal F}$ as~\cite{Gronau:1976ng,Ali:1977eu}
\begin{align}
\frac{d \sigma_{l}}{d^2p_{l\perp} dy_l}
=&
\int
dp_{D\perp}p_{D\perp}dy_D ~
\int d\phi\frac{M_D}{4\pi (p_{D}\cdot p_{l})}
f\left(\frac{ p_{D}\cdot p_{l}}{M_D}\right)
~
\frac{d \sigma_{D}}{d^2p_{D\perp} dy_D}\,
\notag \\
=&
\int
dp_{D\perp}p_{D\perp}dy_D ~{\cal F}(p_{D},p_{l})~
\frac{d \sigma_{D}}{d^2p_{D\perp} dy_D}
\, ,
\label{eq:semileptonic}
\end{align}
where we have defined ${\cal F}$ in the second equality. Here $\phi$ is the azimuthal angle between $p_{D\perp}$ and $p_{l\perp}$, and $p_D \cdot p_l$ is the four-momentum product. The function $f(E_l)$ is the energy distribution of the decay lepton~$l$ with energy $E_l$ in the rest frame of the heavy flavor meson: $f(E_l)=\omega{E_l^2(M_D^2-M_K^2-2M_DE_l)^2}/{(M_D-2E_l)}$ with $M_K=0.497\;{\rm GeV}$ (Kaon's mass) and $M_D=1.86\;{\rm GeV}$ ($D$ meson's mass). We neglect the lepton masses here. The normalization factor reads $\omega = 96/ [(1-8t^2+8t^6-t^8-24t^4\ln t)M_D^6]$ with $t = M_K / M_D$.
We make a comment on the energy distribution of the decay lepton. $f(E_l)$ employed in our calculations is a simple model and was used for describing free quark decay process like $c\rightarrow sl\nu$~\cite{Ali:1977eu}. Nevertheless, as discussed in Ref.~\cite{Ali:1977eu}, the energy distribution of lepton from heavy-flavor meson decay must be similar to that of lepton from free quark decay. Throughout this note, we assume this similarity in order to simplify our calculations.

\subsection{Rapidity dependence}

We incorporate the $x$ dependence or quantum rapidity evolution of the UGDF (\ref{eq:phip}) and the multi-point Wilson line correlator (\ref{eq:phiA}). In the large-$N_c$ limit, these functions are written in terms of the fundamental dipole amplitude $S_x(r_\perp)$, and their rapidity dependence is encoded through $S_x(r_\perp)$. The rapidity evolution of $S_x(r_\perp)$ itself is controlled by the nonlinear Balitsky-Kovchegov equation~\cite{Balitsky:1995ub,Kovchegov:1996ty} with running coupling correction in the evolution kernel (rcBK):
\begin{align}
-\frac{dS_{x}({r_\perp})}{d\ln1/x}
 = \int d^2 r_{1\perp} \mathcal{K}(r_\perp, r_{1\perp}) 
\Big [  S_{x}({r_\perp}) - S_{x}({r_{1\perp}})S_{x}({r_{2\perp}})
\Big ].
\label{eq:bk}
\end{align}
$Y=\ln1/x$ corresponds to the evolution rapidity. Specifically, we use the evolution kernel in Balitsky's prescription~\cite{Balitsky:2006wa} which is given by
\begin{align}
\mathcal{K}(r_\perp,r_{1\perp})=&
\frac{\alpha_s (r^2) N_c} {2\pi^2}\,
\left [
\frac{1}{r_1^2} \left ( \frac{\alpha_s(r_1^2)}{\alpha_s(r_2^2)}-1  \right )
+
\frac{r^2}{r_1^2 r_2^2}
+
\frac{1}{r_2^2} \left ( \frac{\alpha_s(r_2^2)}{\alpha_s(r_1^2)}-1  \right )
\right ]
\label{eq:rcBK-kernel}
\end{align}
with ${r}_\perp= {r}_{1\perp}+ {r}_{2\perp}$ the size of parent dipole prior to one step rapidity evolution. It is well known that global fitting of HERA DIS data below $x_0=0.01$ can constrain the initial condition of the rcBK equation with the following functional form: $S_{{x=x_0}}(r_\perp)=
\exp\left[-\frac{\left(r_\perp^2Q_{s0,p}^2\right)^\gamma}{4}\ln\left(\frac{1}{r_\perp\Lambda}+e\right)\right]$ with $\alpha_s(r^2)= \left [\frac{9}{4\pi} \ln \left (\frac{4 C^2}{r^2 \Lambda^2}+a \right ) \right ]^{-1}$. An example of the fitted parameters set is $Q_{s0,\rm p}^2=0.1597$ GeV, $\gamma=1.118$, $\Lambda=0.241$ GeV, and $C=2.47$. The parameter $a$ is an IR cutoff scale and chosen to satisfy $\alpha_s(r \to \infty)=1.0$. We use this set for describing the proton. For heavy nuclei, we assume a larger value of the initial saturation scale $Q_{s0,A}^2=cA^{1/3}Q_{s0,p}^2$ with $c\lesssim1$~%
  \footnote{
    For phenomenology, $c\sim 0.5$ is favored to reproduce the scattering data
    with nuclear target, such as $J/\psi$ production in pPb collisions
    at the LHC~\cite{Ma:2015sia,Dusling:2009ni,Fujii:2015lld},
    although $c=1$ was a naive anticipation.},
and let the rcBK equation evolve the dipole amplitude in rapidity, which should be reasonable for the minimum bias events in p$A$ collisions. For Pb nucleus, we allow a variation of $Q_{s0,A}^2=(2-4)Q_{s0,p}^2$ in numerical calculations, to estimate the model uncertainty.

At large $x \geq x_0$, we simply adopt the extrapolation ansatz for (\ref{eq:phip})~\cite{Fujii:2006ab,Fujii:2013gxa}: $\varphi_{{\rm p},x}(k_\perp)=\varphi_{{\rm p},x_0}(k_\perp)\left(\frac{1-x}{1-x_0}\right)^4 \left(\frac{x_0}{x}\right)^{0.15}$. We also apply the same procedure to $\phi_{{A},x}({k}_{\perp},{l}_\perp)$. We should discuss the numerical results which are insensitive to this large-$x$ extrapolation.

\section{Remarks on Numerical results}{\label{section:numerical-results}}

In this note, we concentrate on nuclear modification factor which is defined as
\begin{align}
R_{{\rm p}A}=\frac{1}{A}\frac{d^3\sigma_{{\rm p}A}/d^2p_\perp dy}{d^3\sigma_{\rm pp}/d^2p_\perp dy}.
\end{align}
Hereinafter, we regard $p_\perp$ and $y$ as transverse momentum and rapidity of the produced $D$ meson or the decay lepton~$l$. Uncertainties of the cross sections coming from the input parameters including $m_c$, $S_\perp$, $\alpha_s$ and the fragmentation part tend to cancel out by taking the ratio. Thus the CGC framework has a predictive power in describing the nuclear suppression of $D$ meson and $l$ production.

At asymptotically large $p_\perp$ we expect that the particle production occurrs incoherently to give $R_{{\rm p}A}=1$, while our parametrization of the dipole amplitude will yield $R_{{\rm p}A}\sim\frac{1}{A}\frac{\pi R_A^2Q_{s0,A}^{2\gamma}}{\pi R_{\rm p}^2Q_{s0,{\rm p}}^{2\gamma}}$. These are consistent with each other if $\gamma=1$ and $R_A, Q_{s0,A}^2 \propto A^{1/3}$. For $\gamma \ne 1$,  we set $R_A\sim \sqrt{A\frac{Q_{s0,{\rm p}}^{2\gamma}}{Q_{s0,A}^{2\gamma}}}R_{\rm p}$ by requiring $R_{{\rm p}A}=1$ at $p_\perp\rightarrow \infty$ as discussed in Refs.~\cite{Ma:2015sia,Fujii:2015lld}. With $\gamma=1.118$ and $Q_{s0,A}^2=3Q_{s0,{\rm p}}^2$, this constraint gives us $R_A=\sqrt{{A}/{3^\gamma}}R_{\rm p}=7.80R_{\rm p}$.  As for the other initial saturation scale for nucleus, $Q_{s0,A}^2=2Q_{s0,{\rm p}}^2$ and $4Q_{s0,{\rm p}}^2$, it gives $R_A=9.79R_{\rm p}$ and $6.64R_{\rm p}$, respectively. One should note that these effective values of $R_A$ are introduced just to keep the constraint that $R_{{\rm p}A} \to 1$ at large $p_\perp$.

\begin{figure}
	\centering
	\includegraphics[width=8.75cm]{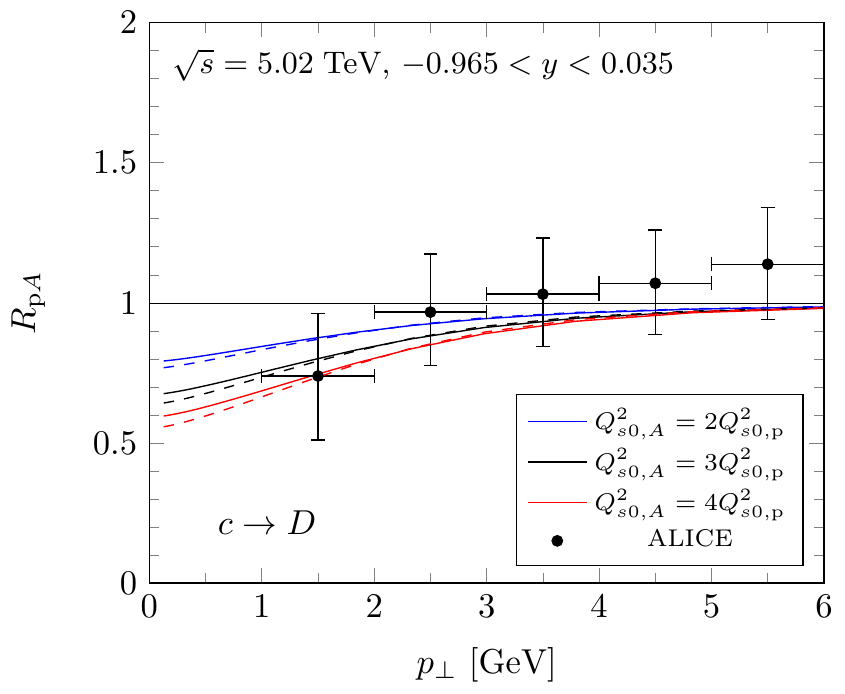}
	\includegraphics[width=8.75cm]{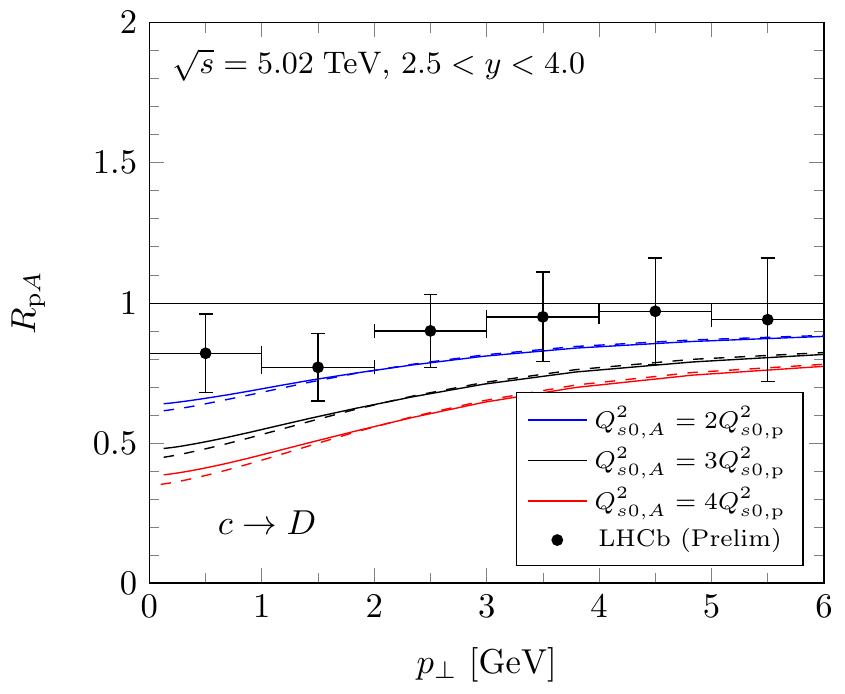}
	\caption{Left : Nuclear modification factor as a function of $p_\perp$ for single $D$ meson production in pPb collisions at the LHC in the mid rapidity region. Blue, black, and red colors reflect the different initial saturation scales as $Q_{s0,A}^2=(2,3,4)Q_{s0,{\rm p}}^2$. Solid (Dashed) lines correspond to the results with $m_c=1.5\;{\rm GeV}$ ($1.2\;{\rm GeV}$). LHC data are taken from Ref.~\cite{Adam:2015qda}. Right : The same plot as the left panel, but at forward rapidity. LHCb preliminary data for $D^0$ production are found in Ref.~\cite{LHCb:2016huj}.
	}
	\label{fig:RpA-pt}
\end{figure}

Let us move to numerical results predicted in the CGC framework at LO. Figure~\ref{fig:RpA-pt} displays $R_{{\rm p}A}$ as a function of $p_\perp$ for $D$　meson production at mid and forward rapidities at the LHC. It also exhibits the dependence of $R_{{\rm p}A}$ on the initial saturation scale of the target nucleus and on the choice of heavy quark mass. The numerical values of $R_{{\rm p}A}(p_\perp)$ are listed in Tables~\ref{tab:D-RpA-seth2}, \ref{tab:D-RpA-seth3}, and \ref{tab:D-RpA-seth4}. In the plot at mid rapidity, the LHC data reported by ALICE Collaboration~\cite{Adam:2015qda} are compared, while the LHCb preliminary data~\cite{LHCb:2016huj} are shown in the plot at forward rapidity. The CGC calculation predicts sizable nuclear suppression of forward $D$ meson production at low $p_\perp$~\footnote{Indeed, similar results are also obtained in Ref.~\cite{Ducloue:2016ywt}, in which optical Glauber model is incorporated for solving the BK equation for heavy nuclei.}. Notice that the numerical values of $R_{{\rm p}A}$ for $D$ meson production corresponds to average values of those for $D^0$, $D^+$, and $D^{\ast+}$ mesons in ALICE data, while LHCb measured $D^0$ meson production only.

\begin{figure}
	\centering
	\includegraphics[width=8.75cm]{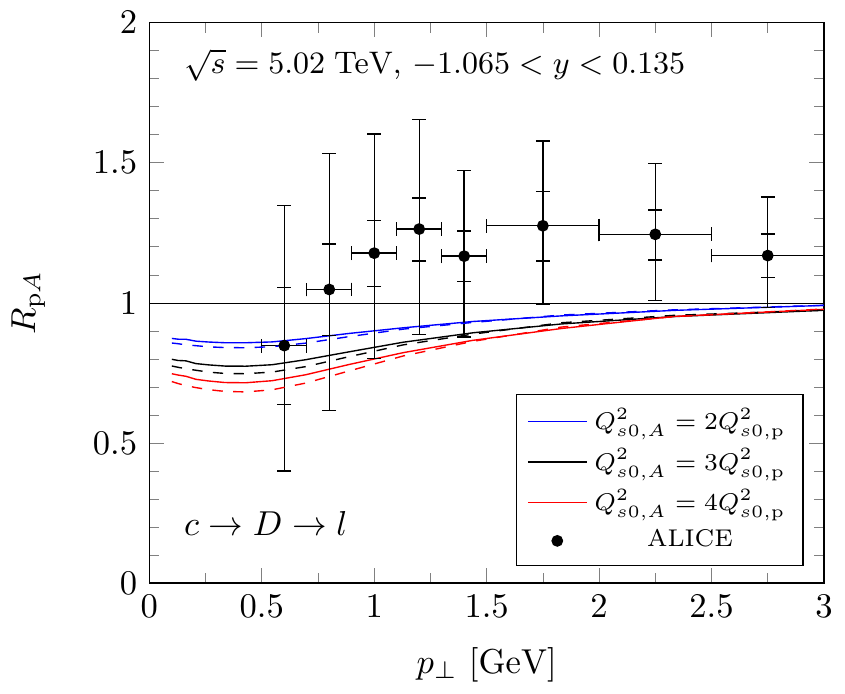}
	\includegraphics[width=8.75cm]{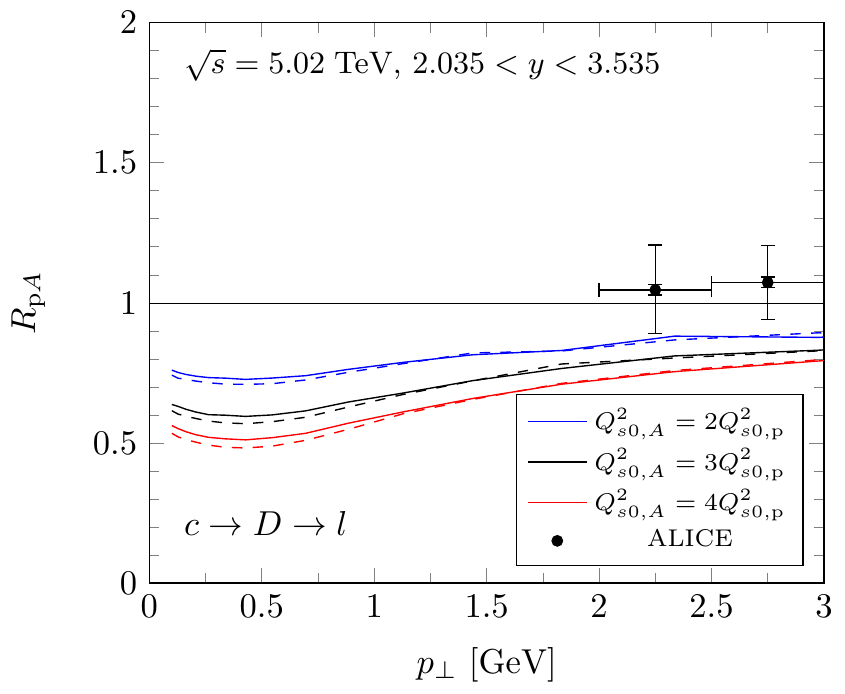}
	\caption{Nuclear modification factor for $l$ production in pPb collisions at the LHC. Notations are the same as in Fig.~\ref{fig:RpA-pt}. LHC data are from Refs.~\cite{Adam:2015qda} and \cite{Acharya:2017hdv}.
	}
	\label{fig:RpA-pt-muon}
\end{figure}

Similarly, we show in Fig.~\ref{fig:RpA-pt-muon} $R_{{\rm p}A}(p_\perp)$ of the produced leptons $l$ from semileptonic decay of $D$ mesons at mid and forward rapidities at the LHC. We allow a variation for the initial nuclear saturation scale and heavy quark mass scale to estimate the model uncertainty. Lepton production in p$A$ collisions at forward rapidity is strongly suppressed at $p_\perp<2\;{\rm GeV}$ as well as $D$ meson production, although further low-$p_\perp$ data are required to examine the suppression quantitatively. The numerical values of $R_{{\rm p}A}$ for the lepton production are listed in Tables~\ref{tab:lep-RpA-seth2}, \ref{tab:lep-RpA-seth3}, and \ref{tab:lep-RpA-seth4}. The LHC data of $e$ production at mid rapidity and $\mu$ production at forward rapidity are taken from Refs.~\cite{Adam:2015qda} and \cite{Acharya:2017hdv}, respectively.

\begin{figure}
	\centering
	\includegraphics[width=8.75cm]{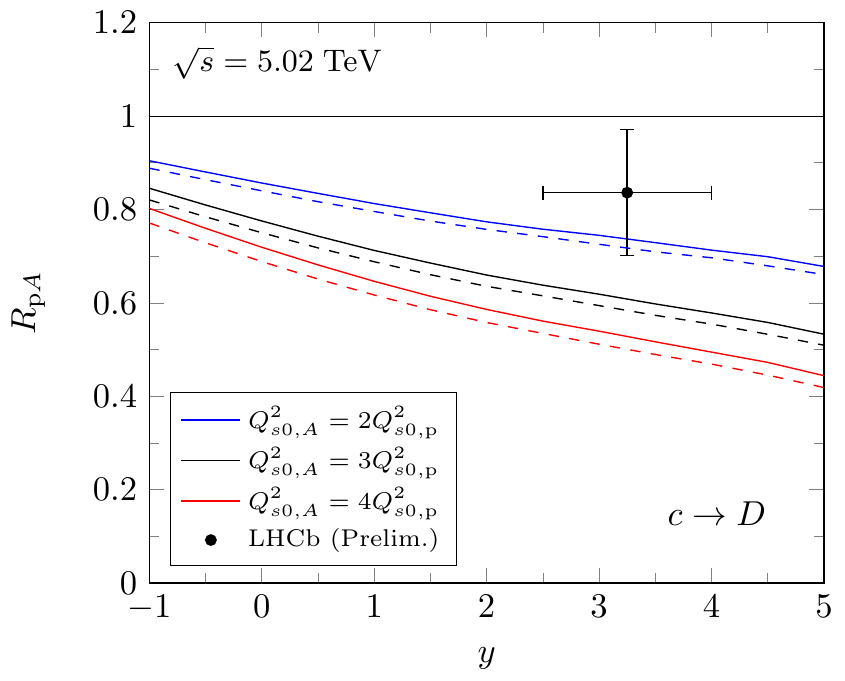}
	\includegraphics[width=8.75cm]{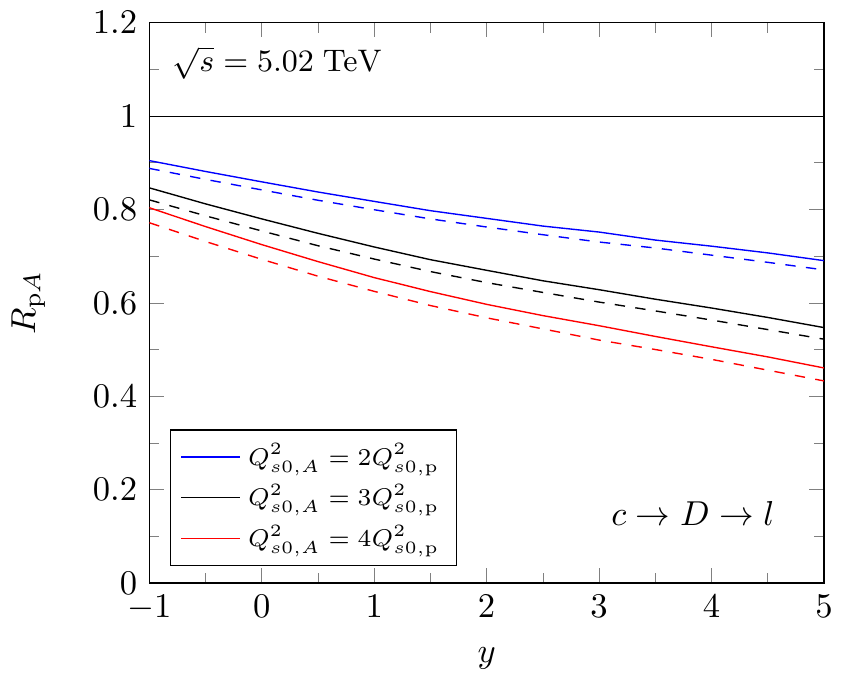}
	\caption{Rapidity dependence of $R_{{\rm p}A}$ at the LHC for $D$ production (left) and $l$ production (right). Dependence on the charm quark mass dependence ($m_c=$ 1.5~(solid) and 1.2~(dashed) GeV) and the initial nuclear saturation scale ($Q_{s0,A}^2/Q_{s0,{\rm p}}^2=$ 2~(blue), 3~(black),  4~(yellow)) are illustrated. In $D$ meson production, the preliminary data from LHCb~\cite{LHCb:2016huj} are compared.
	}
	\label{fig:RpA-y}
\end{figure}

Next, rapidity dependence of $R_{{\rm p}A}$ for $D$ and $l$ production integrated over $p_\perp$ up to $p_\perp=10\; {\rm GeV}$ is shown in Fig.~\ref{fig:RpA-y}. The numerical values of $R_{{\rm p}A}$ vs.\ $y$ for $D$ and $l$ production are given in Tables~\ref{tab:RpA-ydep} and \ref{tab:RpA-ydep-l}, respectively. At present, the preliminary data for $D$ meson production are available from LHCb experiment~\cite{LHCb:2016huj}.  It is very awaited to get new data on rapidity dependence of $R_{{\rm p}A}(y)$ of decay lepton production.

Finally, one must keep in mind the following remarks for interpreting our numerical results:
\begin{itemize}
\item The CGC formula used in our calculation is derived at LO, while the rcBK equation is adopted for describing the rapidity evolution of the dipole amplitude. The full NLO BK equation containing gluon loop corrections in the evolution kernel~\cite{Balitsky:2008zza,Iancu:2015vea} should modify rapidity dependence of the heavy quark production cross section from the rcBK. In addition, higher order perturbative corrections, including the Sudakov logarithms, may be non-negligible. To this end, full NLO CGC formula of heavy quark production in pA collisions is necessary. 
		
	\item In partonic hard scattering at LO, $Q\bar Q$ pair with finite $p_{\perp}$ is likely to be produced in nearly the back-to-back kinematics, where their total momentum ($q_{\rm tot}$) can be much smaller than their relative momentum ($P_{\rm rel}$): $|q_{\rm tot}|\ll |P_{\rm rel}|\sim |M|$ with $M$ being the invariant mass of $Q\bar Q$. This is the very case for the study with transverse momentum dependent (TMD) factorization framework. We naively expect that double logarithmic corrections like $\alpha_s\ln^2\frac{M^2}{q_{\rm tot}^2}\sim{\cal O}(1)$ become important and we can resum them by means of CSS evolution in $b$-space~\cite{Collins:1984kg,Collins:2011zzd}. As is demonstrated in Ref.~\cite{Watanabe:2015yca}, the Sudakov effect can be predominant over (or comparable with) the saturation effect for heavy quark pair production in pp (p$A$) collisions. For single heavy quark production, the phase space of produced antiquark is integrated including the small momentum region. Therefore, implementation of the Sudakov factor in the CGC framework would modify $p_\perp$ spectrum of single heavy quark production.

	\item Our calculations have been restricted to minimum bias events in p$A$ collisions. But centrality dependence is certainly an interesting observable to study the gluon saturation inside a heavy nucleus from experimental data of heavy flavor productions as well as light hadron production. To work on this, it is required to use a consistent approach which describes the nuclear profile and fluctuation effects in p$A$ collisions.~\cite{Albacete:2012xq}.
\end{itemize}

We leave these issues for future study.

\subsection*{Acknowledgement}
We thank T.~Chujo, M.~van~Leeuwen, and T.~Peitzmann for beneficial communications. K.W. is supported by Jefferson Science Associates, LLC under  U.S. DOE Contract \#DE-AC05-06OR23177 and by U.S. DOE Grant \#DE-FG02-97ER41028. Work of HF was supported by Grants-in-Aid for Scientific Research (\#16K05343).



\newpage

\begin{table}[tbp]
	\renewcommand\arraystretch{1.1}
	\begin{center}
		\begin{tabular}{c|c|c||c|c|}
			\cline{2-5}
			&\multicolumn{2}{|c||}{Mid ($-0.965<y<0.035$) }
			&\multicolumn{2}{|c|}{Forward ($2.5<y<4.0$)}\\			
			\hline
			\multicolumn{1}{|c||}{$p_\perp$\;[GeV]} &  \hspace{3pt} $R_{{\rm p}A}$ ($1.5$) \hspace{3pt} & $R_{{\rm p}A}$ ($1.2$) & \hspace{3pt} $R_{{\rm p}A}$ ($1.5$) \hspace{3pt} & $R_{{\rm p}A}$ ($1.2$)\\
			\hline 
			\multicolumn{1}{|c||}{ 0.100 }  &  0.7915  &  0.7674  &  0.6392  & 0.6137   \\
			\multicolumn{1}{|c||}{ 0.127 }  &  0.7925  &  0.7687  &  0.6401  &  0.6151  \\
			\multicolumn{1}{|c||}{ 0.162 }  &  0.7940  &  0.7705  &  0.6413  &  0.6169  \\
			\multicolumn{1}{|c||}{ 0.207 }  &  0.7960  &  0.7730  &  0.6432  &  0.6193   \\
			\multicolumn{1}{|c||}{ 0.264 }  &  0.7987  &  0.7764  &  0.6460  &  0.6227  \\
			\multicolumn{1}{|c||}{ 0.336 }  &  0.8026  &  0.7813  &  0.6499  &  0.6273  \\
			\multicolumn{1}{|c||}{ 0.428 }  &  0.8079  &  0.7879  &  0.6551  &  0.6340  \\
			\multicolumn{1}{|c||}{ 0.546 }  &  0.8151  &  0.7968  &  0.6629  &  0.6434  \\
			\multicolumn{1}{|c||}{ 0.695 }  &  0.8246  &  0.8086  &  0.6717  &  0.6554  \\
			\multicolumn{1}{|c||}{ 0.886 }  &  0.8371  &  0.8239  &  0.6849  &  0.6711  \\
			\multicolumn{1}{|c||}{ 1.129 }  &  0.8528  &  0.8430  &  0.7027  &  0.6918  \\
			\multicolumn{1}{|c||}{ 1.438 }  &  0.8722  &  0.8660  &  0.7244  &  0.7179  \\
			\multicolumn{1}{|c||}{ 1.833 }  &  0.8948  &  0.8925  &  0.7499  &  0.7482  \\
			\multicolumn{1}{|c||}{ 2.336 }  &  0.9194  &  0.9207  &  0.7782  &  0.7806  \\
			\multicolumn{1}{|c||}{ 2.976 }  &  0.9434  &  0.9465  &  0.8096  &  0.8143  \\
			\multicolumn{1}{|c||}{ 3.793 }  &  0.9635  &  0.9658  &  0.8386  &  0.8445  \\
			\multicolumn{1}{|c||}{ 4.833 }  &  0.9776  &  0.9788  &  0.8623  &  0.8678  \\
			\multicolumn{1}{|c||}{ 6.158 }  &  0.9868  &  0.9879  &  0.8828  &  0.8861  \\
			\multicolumn{1}{|c||}{ 7.848 }  &  0.9950  &  0.9935  &  0.9068  &  0.8996  \\
			\multicolumn{1}{|c||}{ 10.00 }  &  1.006  &  0.9997  &  0.9357  &  0.9155  \\
			\hline
		\end{tabular}
		\caption{Numerical values of $R_{{\rm p}A}$ vs $p_\perp$ for $D$ meson production in pPb collisions at $\sqrt{s}=5.02\;{\rm TeV}$ with $Q_{s0,A}^2=2Q_{s0,{\rm p}}^2$. Numbers in the bracket next to $R_{{\rm p}A}$ represent the heavy quark mass value: $m_c=1.5\;$GeV or 1.2\;GeV.
			\label{tab:D-RpA-seth2}}
	\end{center}
\end{table}

\begin{table}[tbp]
	\renewcommand\arraystretch{1.1}
	\begin{center}
		\begin{tabular}{c|c|c||c|c|}
			\cline{2-5}
			&\multicolumn{2}{|c||}{Mid ($-0.965<y<0.035$) }
			&\multicolumn{2}{|c|}{Forward ($2.5<y<4.0$)}\\			
			\hline
			\multicolumn{1}{|c||}{$p_\perp$\;[GeV]} &  \hspace{3pt} $R_{{\rm p}A}$ ($1.5$) \hspace{3pt} & $R_{{\rm p}A}$ ($1.2$) & \hspace{3pt} $R_{{\rm p}A}$ ($1.5$) \hspace{3pt} & $R_{{\rm p}A}$ ($1.2$)\\
			\hline 
\multicolumn{1}{|c||}{ 0.100 }  &  0.6746  &  0.6408  &  0.4795  &  0.4475 \\
\multicolumn{1}{|c||}{ 0.127 }  &  0.6760  &  0.6425  &  0.4805  &  0.4490 \\
\multicolumn{1}{|c||}{ 0.162 }  &  0.6780  &  0.6450  &  0.4819  &  0.4512 \\
\multicolumn{1}{|c||}{ 0.207 }  &  0.6808  &  0.6483  &  0.4840  &  0.4543 \\
\multicolumn{1}{|c||}{ 0.264 }  &  0.6847  &  0.6531  &  0.4873  &  0.4585 \\
\multicolumn{1}{|c||}{ 0.336 }  &  0.6902  &  0.6598  &  0.4921  &  0.4644 \\
\multicolumn{1}{|c||}{ 0.428 }  &  0.6979  &  0.6691  &  0.4988  &  0.4725 \\
\multicolumn{1}{|c||}{ 0.546 }  &  0.7084  &  0.6817  &  0.5078  &  0.4834 \\
\multicolumn{1}{|c||}{ 0.695 }  &  0.7224  &  0.6986  &  0.5203  &  0.4980 \\
\multicolumn{1}{|c||}{ 0.886 }  &  0.7410  &  0.7211  &  0.5373  &  0.5179 \\
\multicolumn{1}{|c||}{ 1.129 }  &  0.7650  &  0.7499  &  0.5601  &  0.5442 \\
\multicolumn{1}{|c||}{ 1.438 }  &  0.7953  &  0.7856  &  0.5888  &  0.5783 \\
\multicolumn{1}{|c||}{ 1.833 }  &  0.8314  &  0.8278  &  0.6234  &  0.6200 \\
\multicolumn{1}{|c||}{ 2.336 }  &  0.8718  &  0.8736  &  0.6643  &  0.6669 \\
\multicolumn{1}{|c||}{ 2.976 }  &  0.9117  &  0.9164  &  0.7100  &  0.7160 \\
\multicolumn{1}{|c||}{ 3.793 }  &  0.9450  &  0.9494  &  0.7533  &  0.7616 \\
\multicolumn{1}{|c||}{ 4.833 }  &  0.9684  &  0.9707  &  0.7891  &  0.7986 \\
\multicolumn{1}{|c||}{ 6.158 }  &  0.9834  &  0.9837  &  0.8197  &  0.8259 \\
\multicolumn{1}{|c||}{ 7.848 }  &  0.9953  &  0.9925  &  0.8479  &  0.8467 \\
\multicolumn{1}{|c||}{ 10.00 }  &  1.011  &  1.003  &  0.8805  &  0.8721 \\
			\hline
		\end{tabular}
		\caption{The same as in Table~\ref{tab:D-RpA-seth2}, but with $Q_{s0,A}^2=3Q_{s0,{\rm p}}^2$.
			\label{tab:D-RpA-seth3}}
	\end{center}
\end{table}

\begin{table}[tbp]
	\renewcommand\arraystretch{1.1}
	\begin{center}
		\begin{tabular}{c|c|c||c|c|}
			\cline{2-5}
			&\multicolumn{2}{|c||}{Mid ($-0.965<y<0.035$) }
			&\multicolumn{2}{|c|}{Forward ($2.5<y<4.0$)}\\			
			\hline
			\multicolumn{1}{|c||}{$p_\perp$\;[GeV]} &  \hspace{3pt} $R_{{\rm p}A}$ ($1.5$) \hspace{3pt} & $R_{{\rm p}A}$ ($1.2$) & \hspace{3pt} $R_{{\rm p}A}$ ($1.5$) \hspace{3pt} & $R_{{\rm p}A}$ ($1.2$)\\
			\hline 
			\multicolumn{1}{|c||}{ 0.100 }  &  0.5945  &  0.5559  &  0.3851  & 0.3524\\
			\multicolumn{1}{|c||}{ 0.127 }  &  0.5962  &  0.5578  &  0.3863  & 0.3537\\
			\multicolumn{1}{|c||}{ 0.162 }  &  0.5984  &  0.5604  &  0.3880  & 0.3558\\
			\multicolumn{1}{|c||}{ 0.207 }  &  0.6016  &  0.5641  &  0.3904  & 0.3588\\
			\multicolumn{1}{|c||}{ 0.264 }  &  0.6061  &  0.5695  &  0.3936  & 0.3629\\
			\multicolumn{1}{|c||}{ 0.336 }  &  0.6124  &  0.5777  &  0.3985  & 0.3689\\
			\multicolumn{1}{|c||}{ 0.428 }  &  0.6213  &  0.5876  &  0.4053  & 0.3772\\
			\multicolumn{1}{|c||}{ 0.546 }  &  0.6336  &  0.6021  &  0.4147  & 0.3885\\
			\multicolumn{1}{|c||}{ 0.695 }  &  0.6502  &  0.6218  &  0.4278  & 0.4040\\
			\multicolumn{1}{|c||}{ 0.886 }  &  0.6724  &  0.6482  &  0.4461  & 0.4253\\
			\multicolumn{1}{|c||}{ 1.129 }   &  0.7015  &  0.6828  &  0.4708  & 0.4540\\
			\multicolumn{1}{|c||}{ 1.438 }  &  0.7389  &  0.7268  &  0.5027  & 0.4917\\
			\multicolumn{1}{|c||}{ 1.833 }  &  0.7846  &  0.7798  &  0.5424  & 0.5391\\
			\multicolumn{1}{|c||}{ 2.336 }  &  0.8367  &  0.8388 &  0.5906  & 0.5934\\
			\multicolumn{1}{|c||}{ 2.976 }  &  0.8897  &  0.8955  &  0.6452  & 0.6513\\
			\multicolumn{1}{|c||}{ 3.793 }  &  0.9344  &  0.9402  &  0.6981  & 0.7068\\
			\multicolumn{1}{|c||}{ 4.833 }  &  0.9648  &  0.9687  &  0.7422  & 0.7516\\
			\multicolumn{1}{|c||}{ 6.158 }  &  0.9826  &  0.9841  &  0.7778  & 0.7847\\
			\multicolumn{1}{|c||}{ 7.848 }  &  0.9938  &  0.9914  &  0.8090  & 0.8062\\
			\multicolumn{1}{|c||}{ 10.00 }  &  1.008  &  1.001  &  0.8454  & 0.8292\\
			\hline
		\end{tabular}
		\caption{The same as in Table~\ref{tab:D-RpA-seth2}, but with $Q_{s0,A}^2=4Q_{s0,{\rm p}}^2$.
			\label{tab:D-RpA-seth4}}
	\end{center}
\end{table}

\begin{table}[tbp]
	\renewcommand\arraystretch{1.1}
	\begin{center}
		\begin{tabular}{c|c|c||c|c|}
			\cline{2-5}
			&\multicolumn{2}{|c||}{Mid ($-1.065<y<0.135$) }
			&\multicolumn{2}{|c|}{Forward ($2.035<y<3.535$)}\\			
			\hline
			\multicolumn{1}{|c||}{$p_\perp$\;[GeV]} &  \hspace{3pt} $R_{{\rm p}A}$ ($1.5$) \hspace{3pt} & $R_{{\rm p}A}$ ($1.2$) & \hspace{3pt} $R_{{\rm p}A}$ ($1.5$) \hspace{3pt} & $R_{{\rm p}A}$ ($1.2$)\\
			\hline 
			\multicolumn{1}{|c||}{ 0.100 }  & 0.8731 & 0.8567 & 0.7600 & 0.7421\\
			\multicolumn{1}{|c||}{ 0.127 }  & 0.8704 & 0.8542 & 0.7516 & 0.7312\\
			\multicolumn{1}{|c||}{ 0.162 }  & 0.8702 & 0.8504 & 0.7446 & 0.7266\\
			\multicolumn{1}{|c||}{ 0.207 }  & 0.8633 & 0.8472 & 0.7384 & 0.7207\\
			\multicolumn{1}{|c||}{ 0.264 }  & 0.8603 & 0.8430 & 0.7332 & 0.7149\\
			\multicolumn{1}{|c||}{ 0.336 }  & 0.8577 & 0.8405 & 0.7314 & 0.7101\\
			\multicolumn{1}{|c||}{ 0.428 }  & 0.8576 & 0.8398 & 0.7269 & 0.7093\\
			\multicolumn{1}{|c||}{ 0.546 }  & 0.8606 & 0.8436 & 0.7317 & 0.7114\\
			\multicolumn{1}{|c||}{ 0.695 }  & 0.8724 & 0.8563 & 0.7401 & 0.7249\\
			\multicolumn{1}{|c||}{ 0.886 }  & 0.8907 & 0.8793 & 0.7628 & 0.7524\\
			\multicolumn{1}{|c||}{ 1.129 }   & 0.9110 & 0.9063 & 0.7877 & 0.7833\\
			\multicolumn{1}{|c||}{ 1.438 }  & 0.9335 & 0.9302 & 0.8148 & 0.8214\\
			\multicolumn{1}{|c||}{ 1.833 }  & 0.9536 & 0.9566 & 0.8300 & 0.8283\\
			\multicolumn{1}{|c||}{ 2.336 }  & 0.9733 & 0.9753 & 0.8813 & 0.8678\\
			\multicolumn{1}{|c||}{ 2.976 }  & 0.9903 & 0.9908 & 0.8769 & 0.8932\\
			\multicolumn{1}{|c||}{ 3.793 }  & 1.022 & 0.9968 & 0.9133 & 0.8767\\
			\hline
		\end{tabular}
		\caption{Numerical values of $R_{{\rm p}A}$ vs $p_\perp$ for $l$ production in pPb collisions at $\sqrt{s}=5.02\;{\rm TeV}$ in mid and forward rapidity regions with use of $Q_{s0,A}^2=2Q_{s0,{\rm p}}^2$. Notations are the same as in Table~\ref{tab:D-RpA-seth2}.
			\label{tab:lep-RpA-seth2}}
	\end{center}
\end{table}

\begin{table}[tbp]
	\renewcommand\arraystretch{1.1}
	\begin{center}
		\begin{tabular}{c|c|c||c|c|}
			\cline{2-5}
			&\multicolumn{2}{|c||}{Mid ($-1.065<y<0.135$) }
			&\multicolumn{2}{|c|}{Forward ($2.035<y<3.535$)}\\			
			\hline
			\multicolumn{1}{|c||}{$p_\perp$\;[GeV]} &  \hspace{3pt} $R_{{\rm p}A}$ ($1.5$) \hspace{3pt} & $R_{{\rm p}A}$ ($1.2$) & \hspace{3pt} $R_{{\rm p}A}$ ($1.5$) \hspace{3pt} & $R_{{\rm p}A}$ ($1.2$)\\
			\hline 
			\multicolumn{1}{|c||}{ 0.100 }  & 0.7986 & 0.7748 & 0.6374 & 0.6151\\
			\multicolumn{1}{|c||}{ 0.127 }  & 0.7945 & 0.7700 & 0.6306 & 0.6032\\
			\multicolumn{1}{|c||}{ 0.162 }  & 0.7931 & 0.7650 & 0.6204 & 0.5955\\
			\multicolumn{1}{|c||}{ 0.207 }  & 0.7833 & 0.7595 & 0.6104 & 0.5873\\
			\multicolumn{1}{|c||}{ 0.264 }  & 0.7787 & 0.7528 & 0.6009 & 0.5781\\
			\multicolumn{1}{|c||}{ 0.336 }  & 0.7745 & 0.7481 & 0.5992 & 0.5718\\
			\multicolumn{1}{|c||}{ 0.428 }  & 0.7740 & 0.7471 & 0.5948 & 0.5687\\
			\multicolumn{1}{|c||}{ 0.546 }  & 0.7792 & 0.7530 & 0.6001 & 0.5759\\
			\multicolumn{1}{|c||}{ 0.695 }  & 0.7972 & 0.7725 & 0.6146 & 0.5917\\
			\multicolumn{1}{|c||}{ 0.886 }  & 0.8249 & 0.8076 & 0.6464 & 0.6284\\
			\multicolumn{1}{|c||}{ 1.129 }   & 0.8596 & 0.8501 & 0.6782 & 0.6732\\
			\multicolumn{1}{|c||}{ 1.438 }  & 0.8931 & 0.8885 & 0.7225 & 0.7217\\
			\multicolumn{1}{|c||}{ 1.833 }  & 0.9248 & 0.9290 & 0.7658 & 0.7818\\
			\multicolumn{1}{|c||}{ 2.336 }  & 0.9512 & 0.9556 & 0.8106 & 0.8032\\
			\multicolumn{1}{|c||}{ 2.976 }  & 0.9727 & 0.9735 & 0.8310 & 0.8287\\
			\multicolumn{1}{|c||}{ 3.793 }  & 1.012 & 0.9782 & 0.8716 & 0.8249\\
			\hline
		\end{tabular}
		\caption{The same as in Table~\ref{tab:lep-RpA-seth2}, but with $Q_{s0,A}^2=3Q_{s0,{\rm p}}^2$.
			\label{tab:lep-RpA-seth3}}
	\end{center}
\end{table}

\begin{table}[tbp]
	\renewcommand\arraystretch{1.1}
	\begin{center}
		\begin{tabular}{c|c|c||c|c|}
			\cline{2-5}
			&\multicolumn{2}{|c||}{Mid\;($-1.065<y<0.135$)}
			&\multicolumn{2}{|c|}{Forward\;($2.035<y<3.535$)}\\			
			\hline
			\multicolumn{1}{|c||}{$p_\perp$\;[GeV]} &  \hspace{3pt} $R_{{\rm p}A}$ ($1.5$) \hspace{3pt} & $R_{{\rm p}A}$ ($1.2$) & \hspace{3pt} $R_{{\rm p}A}$ ($1.5$) \hspace{3pt} & $R_{{\rm p}A}$ ($1.2$)\\
			\hline 
			\multicolumn{1}{|c||}{ 0.100 }  & 0.7472 & 0.7191 & 0.5621 & 0.5345\\
			\multicolumn{1}{|c||}{ 0.127 }  & 0.7424 & 0.7121 & 0.5514 & 0.5218\\
			\multicolumn{1}{|c||}{ 0.162 }  & 0.7381 & 0.7049 & 0.5402 & 0.5120\\
			\multicolumn{1}{|c||}{ 0.207 }  & 0.7272 & 0.6976 & 0.5294 & 0.5028\\
			\multicolumn{1}{|c||}{ 0.264 }  & 0.7212 & 0.6902 & 0.5201 & 0.4924\\
			\multicolumn{1}{|c||}{ 0.336 }  & 0.7157 & 0.6840 & 0.5148 & 0.4846\\
			\multicolumn{1}{|c||}{ 0.428 }  & 0.7149 & 0.6822 & 0.5111 & 0.4822\\
			\multicolumn{1}{|c||}{ 0.546 }  & 0.7222 & 0.6897 & 0.5189 & 0.4886\\
			\multicolumn{1}{|c||}{ 0.695 }  & 0.7435 & 0.7138 & 0.5345 & 0.5095\\
			\multicolumn{1}{|c||}{ 0.886 }  & 0.7796 & 0.7565 & 0.5707 & 0.5490\\
			\multicolumn{1}{|c||}{ 1.129 }   & 0.8223 & 0.8105 & 0.6109 & 0.6043\\
			\multicolumn{1}{|c||}{ 1.438 }  & 0.8662 & 0.8627 & 0.6588 & 0.6555\\
			\multicolumn{1}{|c||}{ 1.833 }  & 0.9086 & 0.9136 & 0.7097 & 0.7126\\
			\multicolumn{1}{|c||}{ 2.336 }  & 0.9518 & 0.9512 & 0.7546 & 0.7585\\
			\multicolumn{1}{|c||}{ 2.976 }  & 0.9760 & 0.9772 & 0.7926 & 0.7968\\
			\multicolumn{1}{|c||}{ 3.793 }  & 1.017 & 0.9832 & 0.8482 & 0.8108\\
			\hline
		\end{tabular}
		\caption{The same as in Table~\ref{tab:lep-RpA-seth2}, but with $Q_{s0,A}^2=4Q_{s0,{\rm p}}^2$.
			\label{tab:lep-RpA-seth4}}
	\end{center}
\end{table}

\begin{table}[tbp]
	\renewcommand\arraystretch{1.1}
	\begin{center}
		\begin{tabular}{c|c|c||c|c||c|c|}
			\cline{2-7}
			&\multicolumn{2}{|c||}{$Q_{s0,A}^2=2Q_{s0,{\rm p}}^2$ }
			&\multicolumn{2}{|c||}{$Q_{s0,A}^2=3Q_{s0,{\rm p}}^2$}			
			&\multicolumn{2}{|c|}{$Q_{s0,A}^2=4Q_{s0,{\rm p}}^2$}\\			
			\hline
			\multicolumn{1}{|c||}{$y$} & \hspace{3pt} $R_{{\rm p}A}$ ($1.5$) \hspace{3pt} & \hspace{3pt} $R_{{\rm p}A}$ ($1.2$) \hspace{3pt} & \hspace{3pt} $R_{{\rm p}A}$ ($1.5$) \hspace{3pt} & \hspace{3pt} $R_{{\rm p}A}$ ($1.2$) \hspace{3pt} & \hspace{3pt} $R_{\rm pA}$ ($1.5$) \hspace{3pt} & \hspace{3pt} $R_{{\rm p}A}$ ($1.2$) \hspace{3pt}\\
			\hline 
			\multicolumn{1}{|c||}{ $-1.0$ }  & 0.9043 & 0.8883 & 0.8454 & 0.8204 & 0.8022 & 0.7710\\
			\multicolumn{1}{|c||}{ $-0.5$ } & 0.8803 & 0.8638 & 0.8096 & 0.7843 & 0.7596 &0.7286 \\
			\multicolumn{1}{|c||}{ $0.0$ }  & 0.8566 & 0.8399 & 0.7754 & 0.7502 & 0.7191 &  0.6884\\
			\multicolumn{1}{|c||}{ $0.5$ }  & 0.8344 & 0.8167 & 0.7428 & 0.7178 & 0.6813 & 0.6511 \\
			\multicolumn{1}{|c||}{ $1.0$ }  & 0.8127 & 0.7957 & 0.7123 & 0.6881 & 0.6461 &  0.6168\\
			\multicolumn{1}{|c||}{ $1.5$ }  & 0.7929 & 0.7752 & 0.6853 & 0.6603 & 0.6142 &  0.5853\\
			\multicolumn{1}{|c||}{ $2.0$ }  & 0.7735 & 0.7573 & 0.6596 & 0.6355 & 0.5860 &  0.5580\\
			\multicolumn{1}{|c||}{ $2.5$ }  & 0.7577 & 0.7418 & 0.6379 & 0.6151 & 0.5609 &  0.5346\\
			\multicolumn{1}{|c||}{ $3.0$ }  & 0.7447 & 0.7255 & 0.6186 & 0.5942 & 0.5394 & 0.5115 \\
			\multicolumn{1}{|c||}{ $3.5$ }  & 0.7290 & 0.7094 & 0.5977 & 0.5733 & 0.5166 & 0.4894 \\
			\multicolumn{1}{|c||}{ $4.0$ }  & 0.7129 & 0.6966 & 0.5784 & 0.5546 & 0.4946 &  0.4689\\
			\multicolumn{1}{|c||}{ $4.5$ }  & 0.6988 & 0.6793 & 0.5581 & 0.5328 & 0.4725 &  0.4453\\
			\multicolumn{1}{|c||}{ $5.0$ }  & 0.6780 & 0.6607 & 0.5329 & 0.5094 & 0.4441 & 0.4186 \\
			\hline
		\end{tabular}
		\caption{ Rapidity dependence of $R_{{\rm p}A}$ for $D$ meson production in pPb collisions at $\sqrt s=5.02\;{\rm TeV}$ with several values of the initial saturation scale for target nucleus.
			\label{tab:RpA-ydep}}
	\end{center}
\end{table}

\begin{table}[tbp]
	\renewcommand\arraystretch{1.1}
	\begin{center}
		\begin{tabular}{c|c|c||c|c||c|c|}
			\cline{2-7}
			&\multicolumn{2}{|c||}{$Q_{s0,A}^2=2Q_{s0,{\rm p}}^2$ }
			&\multicolumn{2}{|c||}{$Q_{s0,A}^2=3Q_{s0,{\rm p}}^2$}			
			&\multicolumn{2}{|c|}{$Q_{s0,A}^2=4Q_{s0,{\rm p}}^2$}\\			
			\hline
			\multicolumn{1}{|c||}{$y$} & \hspace{3pt} $R_{{\rm p}A}$ ($1.5$) \hspace{3pt} & \hspace{3pt} $R_{{\rm p}A}$ ($1.2$) \hspace{3pt} & \hspace{3pt} $R_{{\rm p}A}$ ($1.5$) \hspace{3pt} & \hspace{3pt} $R_{{\rm p}A}$ ($1.2$) \hspace{3pt} & \hspace{3pt} $R_{{\rm p}A}$ ($1.5$) \hspace{3pt} & \hspace{3pt} $R_{{\rm p}A}$ ($1.2$) \hspace{3pt}\\
			\hline 
			\multicolumn{1}{|c||}{ $-1.0$ }  & 0.9046 & 0.8879 & 0.8462 & 0.8206 & 0.8038 & 0.7716\\
			\multicolumn{1}{|c||}{ $-0.5$ } & 0.8814 & 0.8644 & 0.8119 & 0.7864 & 0.7633 & 0.7314\\
			\multicolumn{1}{|c||}{ $0.0$ }  & 0.8591 & 0.8418 & 0.7798 & 0.7540 & 0.7247 & 0.6932\\
			\multicolumn{1}{|c||}{ $0.5$ }  &0.8374  & 0.8197 & 0.7489 & 0.7223 & 0.6884 & 0.6571\\
			\multicolumn{1}{|c||}{ $1.0$ }  & 0.8173 & 0.7994 & 0.7197 & 0.6938 & 0.6541 & 0.6247\\
			\multicolumn{1}{|c||}{ $1.5$ }  & 0.7974 & 0.7798 & 0.6925 & 0.6670 & 0.6241 & 0.5943\\
			\multicolumn{1}{|c||}{ $2.0$ }  & 0.7809 & 0.7624 & 0.6697 & 0.6434 & 0.5968 & 0.5680\\
			\multicolumn{1}{|c||}{ $2.5$ }  & 0.7644 & 0.7460 & 0.6473 & 0.6224 & 0.5728 & 0.5443\\
			\multicolumn{1}{|c||}{ $3.0$ }  & 0.7516 & 0.7303 & 0.6282 & 0.6017 & 0.5511 & 0.5203\\
			\multicolumn{1}{|c||}{ $3.5$ }  & 0.7345 & 0.7175 & 0.6079 & 0.5828 & 0.5282 & 0.5001\\
			\multicolumn{1}{|c||}{ $4.0$ }  & 0.7214 & 0.7022 & 0.5890 & 0.5632 & 0.5060 & 0.4791\\
			\multicolumn{1}{|c||}{ $4.5$ }  & 0.7070 & 0.6868 & 0.5688 & 0.5428 & 0.4843 & 0.4559\\
			\multicolumn{1}{|c||}{ $5.0$ }  & 0.6904 & 0.6706 & 0.5471 & 0.5222 & 0.4606 & 0.4329\\
			\hline
		\end{tabular}
		\caption{$R_{{\rm p}A}$ vs $y$ for $l$ production in pPb collisions at the LHC. Notations are the same as in Table~\ref{tab:RpA-ydep}.
			\label{tab:RpA-ydep-l}}
	\end{center}
\end{table}

\end{document}